\documentclass[floatfix,aps,pre,showpacs]{revtex4}
\pdfoutput=1
\usepackage{graphicx}
\usepackage{amsmath}
\usepackage{amsfonts}
\usepackage{amssymb}
\begin{document}

\title{A message-passing scheme for non-equilibrium stationary states}

\author{Erik Aurell}
\affiliation{Department of Computational Biology, AlbaNova University Centre, 106 91 Stockholm, Sweden}
\altaffiliation{Department of Information and Computer Science, Aalto University, Finland}
\altaffiliation{ACCESS Linnaeus Centre, KTH - Royal Institute of Technology, Stockholm, Sweden}
\author{Hamed Mahmoudi}
\affiliation{Department of Information and Computer Science, Aalto University, Finland}

\date{\today}

\begin{abstract}
We study stationary states in a diluted 
asymmetric (kinetic) Ising model. 
We apply the recently introduced dynamic cavity 
method to compute magnetizations of these
stationary states.
Depending on the update rule, different versions
of the dynamic cavity method apply. We here study
synchronous updates and random sequential updates,
and compare local 
properties computed by the dynamic cavity method to 
numerical simulations.
Using both types of updates, the dynamic cavity method is highly
accurate at high enough temperatures.
At low enough temperatures, 
for sequential updates the dynamic cavity method 
tends to a fixed point, but which does not agree with 
numerical simulations, while for parallel updates, the
dynamic cavity method may display oscillatory behavior.
When it converges and is accurate, the dynamic
cavity method offers a huge speed-up compared to Monte Carlo,
particularly for large systems.  
\end{abstract}


\maketitle 

\section{`Introduction}
Stationary states of classical equilibrium systems are
described by the Boltzmann-Gibbs measure. 
In complex systems, the exact computation of even
local properties (marginals) is not feasible, and
perturbative methods or other approximations therefore
have to be used. Much attention has over the last
decade been given to the Belief Propagation (BP) 
or Bethe-Peierls ansatz
class of approximations of the marginals, which are
exact if the underlying graph
of interactions is a tree, and generally expected to be accurate
if the underlying graph is locally tree-like~\cite{mezard2,yedidia}.

In contrast to equilibrium systems, 
non-equilibrium systems do not admit a similar, universal, 
description of their stationary states. We here take a
small step in this direction. We show that the recently
introduced dynamic cavity method, essentially a Bethe-Peierls
ansatz on spin system histories, can be used effectively 
to compute local properties in stationary states
when the underlying graph of interactions is sparse, and 
locally tree-like. A key technical step to make this a
computationally feasible scheme is a stationarity assumption,
here termed {\it time factorization}.
With this step we find a message-passing scheme similar
to Belief Propagation, but for dynamics, and for 
non-equilibrium systems. When these approximations work, at
sufficiently high temperatures (weak interactions) they
offer a huge speed-up compared to Monte Carlo

We now give a synthetic overview of the dynamic cavity approach. A total history
of a collection of spins, evolving in discrete time steps according to some dynamics,
can be described by the total joint probability $p(\vec{\sigma}(0)\ldots\vec{\sigma}(t))$.
From such a total joint probability one can construct the marginal probability
over the history of one spin $p_i(\sigma_i(0)\ldots\sigma_i(t))$. If the underlying 
graph of interactions is locally tree-like, then this marginal probability can
be expressed using "messages" from the neighboring nodes of the type
$\mu_{j\to i}(\sigma_j(0)\ldots\sigma_j(t))$, and these messages in turn obey
update rules similar (in principle) to Belief Propagation. In contrast to 
cavity method in equilibrium analysis, messages carry the whole information of
spin histories over time. The difficulty arises
when we want to marginalize further to the configuration of one spin at one
time in its history. In general, the resulting equations are non-Markovian, and
hard to solve. One level of approximation is to assume that the messages
factorize in time, as $\mu_{j\to i}(\sigma_j(0)\ldots\sigma_j(t))=\prod_{s=0}^t \mu^{s}_{j\to i}(\sigma_j(s))$,
and in stationary state we can further assume that the terms in the product
($\mu^{(s)}_{j\to i}(\sigma_j(s))$  are all the same function \textit{i.e.} do not depend on $s$).
The terms in the factorized messages ($\mu_{j\to i}(\sigma_j)$) then obey
a set of distributed equations analogous to (but more complex than) Belief Propagation.
Depending on the dynamics of the spin system, the resulting equations 
will differ. We here investigate both parallel
updates, as studied recently by Neri and Boll\'e~\cite{bolle},
as well as random sequential updates.

A summary of our results is as follows. We study stationary states in a diluted 
asymmetric (kinetic) Ising model. 
This well-studied non-equilibrium model
has (qualitatively) three parameters: the asymmetry degree (how much
the system is out of equilibrium); the connectivity of the underlying interaction graph
(how accurate a Bethe-Peierls ansatz can be expected to be), and the strength
of the interactions, customarily denoted inverse "temperature".
Furthermore, the type of the dynamics influences behavior. We here
study parallel (synchronous) updates and sequential updates (one spin
at a time).
We compare dynamic
cavity results to numerical simulations of the spin system 
dynamics, which we will refer to as Monte Carlo.
The first result is that the more asymmetric the network, the better dynamic cavity method agrees with Monte Carlo.
This can be understood as an effect of lack of memory in the graph: message go out, and rarely come back.
The second result is that the sparser the underlying graph of interactions, the better dynamic cavity method
agrees with Monte Carlo. This can be understood as an effect that the Bethe-Peierls approximation in
itself being more accurate for locally tree-like graphs. The third result is that dynamic cavity agrees well with
Monte Carlo at high temperature, but deviates from numerical simulations of the full dynamics at low
enough temperatures. The way in which dynamic cavity diverges from Monte Carlo at low temperature
depends on the update rule. For random sequential dynamics we find that dynamic cavity (in the time factorized
approximation for the stationary state) goes to a fixed point, but which does not agree with the stationary
state estimated from Monte Carlo. For parallel updates on the other hand, we find that dynamic cavity
method at low enough temperature does not go to a fixed point. As the system size increases, we find
that dynamic cavity matches Monte Carlo better, and that the fluctuations in parallel updates diminish. 
In summary, the parameter ranges where dynamic cavity method agrees well with Monte Carlo
seem to be rather wide, and not decreasing with system size. We therefore believe that dynamic
cavity as presented here will be a promising avenue to compute and explore stationary states
of large non-equilibrium systems.

The paper is organized as follows. In section(\ref{sec:model}) we 
briefly describe the model, introduce the macroscopic observables 
and we study the dynamic update rules. Section(\ref{sec:naive_MF})
summarizes previous studies using ``naive'' mean field theory for the 
 kinetic Ising model. In section(\ref{sec:BP}) 
we introduce the dynamic cavity method for parallel and sequential
updates, and in section(\ref{sec:result}) we compare dynamic cavity
to direct numerical simulation (Monte Carlo).
Section(\ref{sec:conclusion}) contains concluding remarks.

\section{The diluted asymmetric (kinetic) spin glass}
\label{sec:model}
Kinetic Ising models
were originally motivated by neural networks, to extend
the Hopfield model to asymmetric interactions~\cite{hopfield}. 
These non-equilibrium systems with random interactions have formal
similarities to the equilibrium and especially
out-of-equilibrium (relaxation)
dynamics of spin glasses, and therefore have a long
history of study using methods from that field.
Sompolinsky and Zippelius~\cite{sompol2} introduced 
a formalism based on the Langvin equations 
of spherical spin models.
An analogous approach was then proposed by Sommers in \cite{sommers}
through a path integral analysis of the Glauber dynamics. 
More recently, 
the dynamic replica theory has been developed,
partly with an application to this kind of systems
in mind~\cite{coolen1,mozeika1,mozeika2}.
A general feature of dynamic replica theory is an average
over disorder (average over a class of random graphs and random
interactions); in addition technical assumptions may be needed
such as considering the dynamics of some finite set of appropriate observables
\cite{coolen1}. In~\cite{hatchett} a combination of dynamic replica theory and the cavity method (equilibrium) concept was applied to finitely
connected disordered spin systems.
An alternative approach to dynamic replica theory is generating functional analysis (GFA). 
The GFA has a long history in analyzing non-equilibrium statistical mechanics of disordered systems~\cite{dominics}. 
In particular, it allows us to study systems with non symmetric interacting couplings~\cite{sompol1, coolen2}.
More recently it has been
developed to study the dynamics of spin glass systems with finite connectivity interactions~\cite{coolen2}.
In its general formalism, GFA aims at solving the dynamics of spin system exactly, however,
due to the complicated nature of problem, one needs 
either to use a perturbative approach~\cite{kiemes} or to restrict the analysis of GFA up to some approximation levels~\cite{mimura,coolen2}.
For ferromagnetic systems with regular connectivity, which is much simpler than spin glasses,
a recursive set of dynamical equations can be derived for some finite macroscopic observations~\cite{martin2}.
 
The dynamic cavity method shares some of the features of the dynamic
replica theory and generating functional analysis. In equilibrium analysis, 
the Ising spin glass systems in Bethe lattice model have been solved by cavity method on the level of replica symmetry breaking. 
However in contrast to replica method, cavity applies to one single graph instance (one
single set of interactions). Neri and Boll\'e~\cite{bolle} and Kanoria
and Montanari~\cite{montanari} considered dynamic cavity method in parallel updates under respectively
Glauber dynamics and majority rule dynamics.
In the present work we extend the dynamic cavity method to random
sequential updates, and investigate stationary states of the diluted 
asymmetric spin glass over the wide range of parameters, for both
parallel updates and random sequential updates. 

The asymmetric diluted Ising model is defined over 
a set of $N$ binary variables $\vec{\sigma}=\{\sigma_1,\ldots,\sigma_N\}$,
and an asymmetric graph $G=(V,E)$ where
$V$ is a set of $N$ vertices, and $E$ is a set of directed edges.
To each vertex $v_i$ is associated a binary variable $\sigma_i$.
The graphs $G$ are taken from random graph ensembles with
bounded average connectivity. Following the parameterization of~\cite{coolen2}
we introduce a connectivity matrix $c_{ij}$, where $c_{ij}=1$ if there is
a link from vertex $i$ to vertex $j$, $c_{ij}=0$ otherwise, and matrix elements
$c_{ij}$ and $c_{kl}$ are independent unless $\{kl\}=\{ji\}$.  
The random graph is specified by marginal (one-link) distributions
\begin{equation}
p(c_{ij}) = \frac{c}{N}\delta_{1,c_{ij}} + (1 - \frac{c}{N})\, \delta_{0,c_{ij}}\,\,\,.
\label{eq:connec}
 \end{equation}   
and the conditional distributions
\begin{equation}
p(c_{ij}\,|\,c_{ji}) = \epsilon \delta_{c_{ij},c_{ji}} + (1-\epsilon) \, p(c_{ij})\,\,\,.
\label{eq:asym}
\end{equation} 
In this model the average degree distribution is given by $c$,
and the asymmetry is controlled by $\epsilon\in[\,0,1\,]$.
The two extreme values of $\epsilon$ give respectively an asymmetric network 
($\epsilon=0$), where the probabilities of having two directed links
between pairs of variables are uncorrelated, and the fully symmetric network ($\epsilon=1$)
where the two links $i\to j$ and $j\to i$ are present or absent together. 
The parameter set is completed by a (real-valued) interaction matrix
$\frac{J_{ij}}{c}$. We will always take $J_{ij}$ 
to be independent identically distributed random variables with 
zero mean and unit variance (Gaussian or binary) such that for the fully
connected networks $(c=N)$, the the interactions scale as the 
Sherrington-Kirkpatrick model. 

The definition of the kinetic Ising model is completed by a dynamics,
or a spin update rule. In the synchronous update rules, which will be considered here,
at each (discrete) time,
a set of candidate spins are selected, and then updated according to the rule
\begin{eqnarray}
\sigma_i(t+\Delta t) = \left\{
\begin{array}{c cl}
+1 &{\rm with \,\,probability} \,\,\,\, \{1+\exp(2\beta\,h_i(t+\Delta t))\}^{-1}&\\
-1  &{\rm with \,\,probability} \,\,\,\, \{1+\exp(-2\beta\,h_i(t+\Delta t))\}^{-1}& 
\end{array}
\right.
\label{eq:dynamic}
\end{eqnarray}
where $\Delta t$ is the time interval in which the update procedure 
takes place and $h_i(t)$ is the effective field 
on spin $i$ at time step $t$
\begin{equation}
h_i(t) = \sum_{j\in\partial i} \, \frac{J_{ij}}{c}\,\sigma_j(t-\Delta t)+ \theta_i(t) \,\,\,\,\, . 
\label{eq:eff_field}
\end{equation}
and the parameter $\beta$, analogous to inverse temperature, is
a measure of the overall strength of the interactions.
The notation $j \in \partial i$ in (\ref{eq:dynamic}) and
(\ref{eq:eff_field})
indicates all vertices having a direct
links to node $i$ (defined by $c_{ji}=1$) and $\theta_i$ is the
(possibly time-dependent) external field acting on spin $i$.

We will consider two cases of synchronous updates: either
all spins are selected and updated at each time step, or only
one spin is randomly selected and updated in each time step.
We refer to the first update rule as   
{\it parallel}, and the second as {\it sequential}.
The time interval between updates is taken 
$\Delta t =1$ in parallel updates, and
$\Delta t= \frac{1}{N}$ in sequential updates,
such that in both cases ${\cal O}(N)$ spins are
updated per unit time.

The joint probability distribution over all the spin histories
$p(\vec{\sigma}(0),\ldots,\vec{\sigma}(t))$ has in principle the following simple
Markovian form
\begin{equation}
p(\vec{\sigma}(0)\ldots,\vec{\sigma}(t)) = \prod_{s=1}^t \,W[\,\vec{\sigma}(s)\,|\,\vec{h}(s)]\,p(\vec{\sigma}(0))
\label{eq:prob_dyn}
\end{equation} 
where $W$ is the appropriate transition matrix describing dynamics and 
updates. 
Solving equation~(\ref{eq:prob_dyn}) is in general infeasible for large system size, since it 
consists of $2^{t}\,2^{N}$ equations corresponding to all possible spin history configurations.
One therefore needs to restrict the analysis to some restricted set of observables.
In this paper we are interested in approximations built on marginal
probabilities.
The evolution of a a single spin is (trivially) defined by 
summing over the histories of all spins except one
\begin{equation}
p_i(\sigma_i(0),\ldots,\sigma_i(t)) = \sum_{\vec{\sigma}_{\setminus i}(0),\ldots,\vec{\sigma}_{\setminus i}(t)} \, p(\vec{\sigma}(0),\ldots,\vec{\sigma}(t))
\label{eq:marginal}
\end{equation}
and similarly  for pairwise joint probability of the histories of two spins ($t > t'$)
\begin{equation}
p_{ij}(\sigma_i(0),\ldots\sigma_i(t),\sigma_j(0),\ldots,\sigma_j(t')) = \sum_{\vec{\sigma}_{\setminus i,j}(0),\ldots,\vec{\sigma}_{\setminus i,j}(t)}\, p(\vec{\sigma}(0),\ldots,\vec{\sigma}(t)) \,\,\,\,\, .
\label{eq:correlation}
\end{equation}
Consequently, the time evolution of single site magnetization and the pairwise correlations can be obtained from Eq(\ref{eq:marginal}) and Eq(\ref{eq:correlation})
\begin{eqnarray}
\label{eq:marg_prob}
p_i(\sigma_i(t)) &=& \frac{1+m_i(t)\,s_i(t)}{2} \\
p_{ij}(\sigma_i(t),\sigma_j(t')) &=& \left[1+m_i(t)\sigma_i(t)+m_j(t')\sigma_j(t')+c_{ij}(t,t')\sigma_i(t)\,\sigma_j(t')\right]/4
\end{eqnarray}
Computing the marginal probabilities directly is clearly an intractable
 problem for large system sizes, since the exact enumeration requires 
 summation over an exponentially large number of states. 
 The situation is even more involved for time dependent quantities
  since in addition the effect of past history must be taken into account.

\section{Naive mean-field approximation}
\label{sec:naive_MF}
The dynamics of spin glass models has been widely studied using 
mean-field approximations; we follow recent practice in referring to 
this level  of approximation as naive mean-field~\cite{erik}. 
In~\cite{derrida,sompol1} such a theory
was proposed to describe the dynamics of Little-Hopfield model 
in the case of fully asymmetric networks. 
Here we briefly summarize the naive mean field theory for diluted asymmetric 
systems.
The time evolution of magnetization and correlations for the parallel update
defined in Eq(\ref{eq:prob_dyn}) can be explicitly written as
\begin{equation}
m_i(t) = \left<\tanh(\beta\,h_i(t))\right>
\label{eq:exact_parallel}
\end{equation}
where $\left<\ldots\right>$ represents the average over probability distribution at time $t$. 
Similarly for the sequential update we have
\begin{equation}
m_i(t+\frac{1}{N}) = (1-\frac{1}{N})m_i(t) + \frac{1}{N}\,\left<\tanh(\beta\,h_i(t))\right>
\label{eq:exact_sequen}
\end{equation}
Equations (\ref{eq:exact_parallel} and (\ref{eq:exact_sequen})
are yet exact. The right hand side of equations is however not easy to compute.
The assumption in naive mean field theory is to
substitute the effective field $h_i(t)$ at time $t$ with a time dependent 
random Gaussian noise which does not contain spins configurations~\cite{sompol1}.
The formula for the time evolution of the magnetization (expectation value of
a single spin) is then
\begin{eqnarray}
m_i(t+1) &=& \tanh\left(\beta \hat{h}_i(t)\right) \,\,\,\,\,\,\,\,\,\,\,\,\,\,\,\,\,\,\,\,\, \,\,\,\,\,\,\, \,\,\,\,\,\,\, \,\,\,\,\,\,\, \,\,\,\,\,\,\, \,\,\,\,\,\,\, \,\,\,\,\,\,\,   {\rm parallel \,\,update}\\
m_i(t+\frac{1}{N}) &=& m_i(t) + \frac{1}{N}\left( \tanh(\beta\,\hat{h}_i(t))-m_i(t)\right) \,\,\,\,\,\,\,  {\rm sequential \,\,update}
\end{eqnarray}
where $\hat{h}_i(t)$ is given by the mean values of 
spins neighboring $i$ and a Gaussian noise reflecting the 
effect of neighbors on the dynamic of spin $i$ (see~\cite{sompol1}and~\cite{sommers}).
The fixed point of two dynamic updates coincide, but system size 
is important for the stability of fixed point in the sequential 
update whereas it has no effect on the parallel update scheme~\cite{parisi1,wang}.
The naive mean-field approximation, although introduced 
quite some time ago, remains
a main theoretical tool to analyze kinetic Ising models. 
More recently, these approximations have been used 
as the basis for ``kinetic Ising'' reconstruction 
schemes~\cite{hongli1,yasserhertz,hongli2}. 

\section{The dynamic cavity method} 
\label{sec:BP}
In this paper we use the terms Bethe-Peierls approximation (BP) and cavity method
interchangeably. Their modern use grow out of replica theory in spin glasses,
but in a form which can be applied to a single instance~\cite{mezard1,mezard2,martin}. 
We use the term \textit{dynamic cavity method} for the use of the
Bethe-Peierls approximation on spin histories.

\begin{figure}[htb]
\vspace{0.6cm}
\includegraphics[width=0.45\columnwidth]{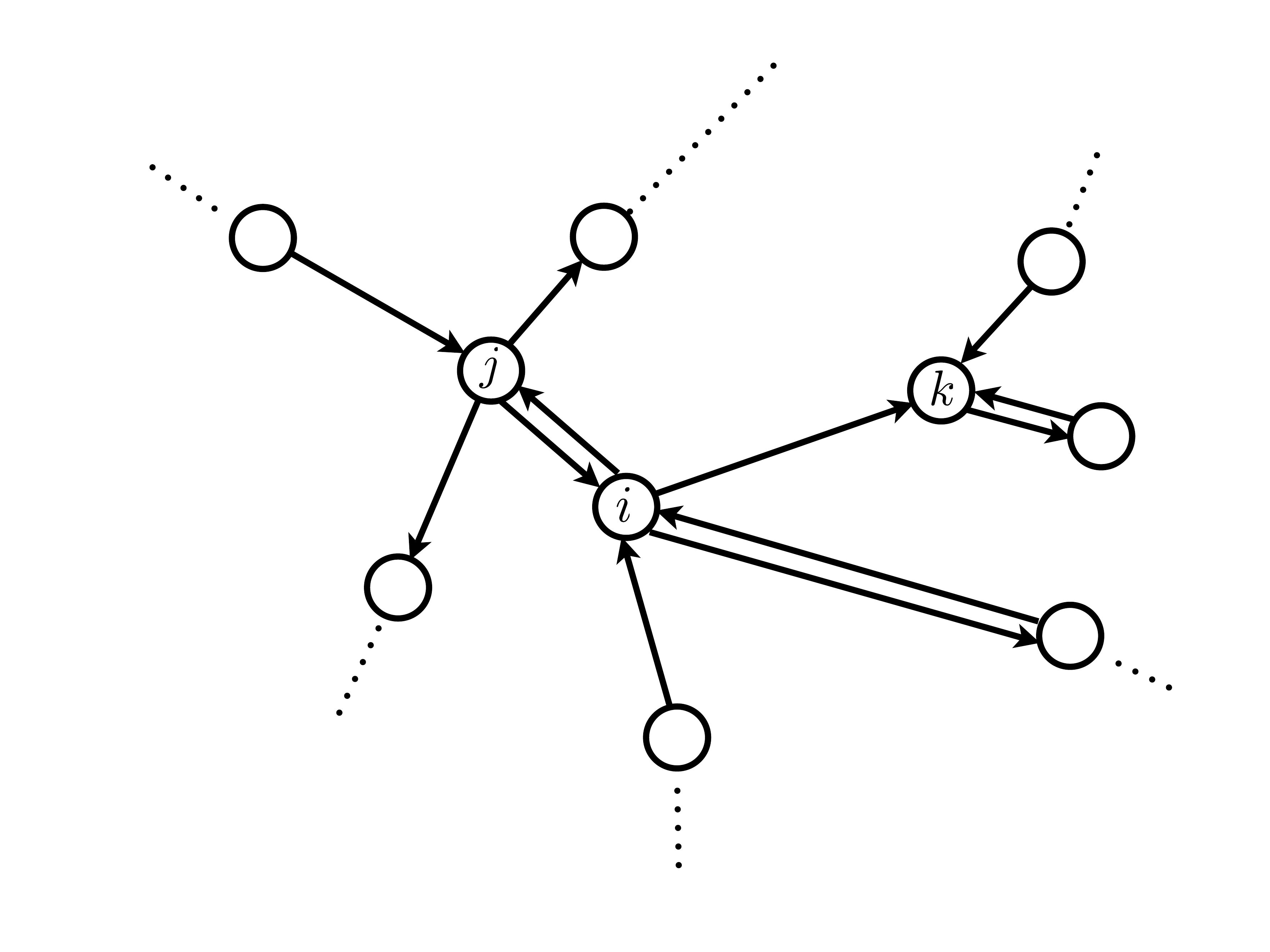}
\hspace{3mm}
\includegraphics[width=0.45\columnwidth]{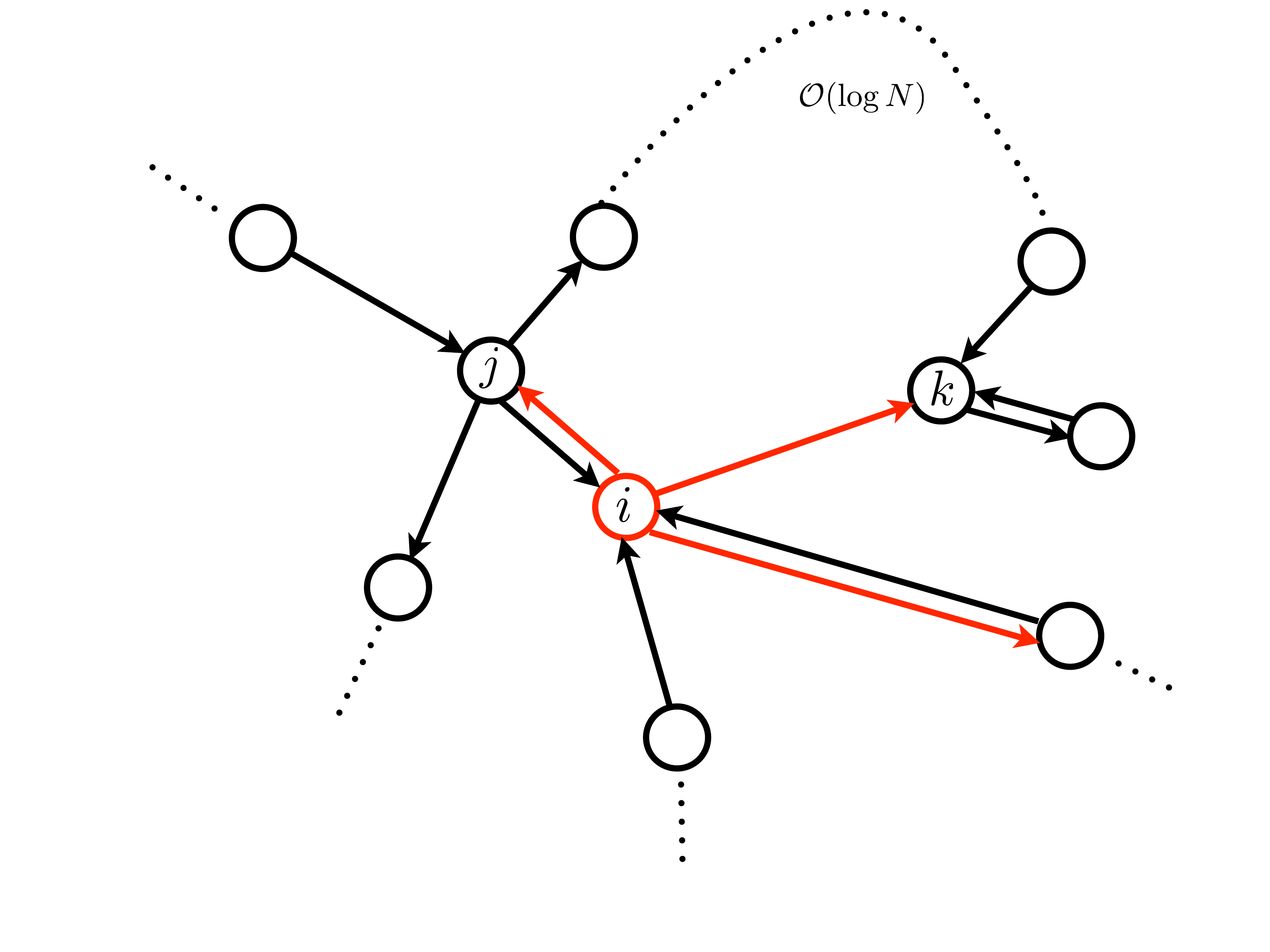}
\caption{Left figure: a directed network representation
 of the asymmetric Ising model. Interacting pairs are 
 connected by directed edges where the arrows indicate the 
the asymmetric nature of model. Right figure: the cavity graph 
created by fixing the spin $i$
to have value $s_i$. This  effects other spins 
which were connected to vertex $i$ by an incoming edge from $i$ (shown by red color). 
In graphs with tree structure, this leads immediately to a factorized 
probability distribution for the set of spins in 
different branches outgoing from node $i$.}
\label{fig:cavity}
\end{figure}

The main idea of BP is to ignore long loop correlations, since BP
is exact on trees.
BP can then be described by a set of auxiliary graphs called
cavity graphs, which are identical to the original 
graphs but with one of the vertices and its associated
variable removed. 
The effects of removing a target node $i$ appear in the effective
 fields Eq(\ref{eq:eff_field}) acting on the set of variables 
neighboring  node $i$ with direct interactions outgoing from spin $i$.
Fig\ref{fig:cavity} illustrates the argument graphically.
By the assumption that short loops are absent and
long loops are ignored, the variables associated
to the vertices which were connected to the removed vertex
(denoted by $i$ in Fig\ref{fig:cavity}) 
are independent in the cavity graph.
We can then consider the joint probability of the histories of
all the other spins
$p^{(i)}(\vec{\sigma}_{\setminus i}(0),\ldots,\vec{\sigma}_{\setminus i}(t)\,| \vec{\theta}^{(i)}_{\setminus i}(0),\ldots,\vec{\theta}^{(i)}_{\setminus i}(t))$
under the influence of the external fields modified by the action of spin $i$
\textit{i.e.}
$\theta^{(i)}_{j}(s)=\theta_{j}(s)+J_{ji}\sigma_i(s-\Delta t)$. 
Since the removed vertex only modifies the effective fields of
its neighbors with an outgoing edge
we can further marginalize the joint probability distribution
 to the set $\partial i$ of spins ``in the cavity'' which were
directly connected to vertex $i$,
$p^{(i)}(\sigma_{\partial i}(0),\ldots,\sigma_{\partial i}(t)\,| \theta^{(i)}_{\partial i}(0),\ldots,
\theta^{(i)}_{\partial i}(t))$. 
The assumption of no loops means that the set of spins neighboring vertex $i$ ``in the cavity'' produced
by removing vertex $i$ are independent,
and that therefore this joint probability over these spin histories is factorized as 
\begin{equation}
p^{(i)}(\sigma_{\partial i}(0),\ldots,\sigma_{\partial i}(t)\,| \theta^{(i)}_{\partial i}(0),\ldots,\theta^{(i)}_{\partial i}(t)) =  \prod_{j\in \partial i}  \,\, 
\mu_{j\to i}(\sigma_j(0),\ldots,\sigma_j(t)|\theta^{(i)}_{j}(0),\ldots,\theta^{(i)}_{j}(t)) 
\label{eq:fact}
\end{equation} 
The above approximation is exact in trees, but in general 
it can be only considered as an approximation. 
For random graph ensembles with diluted interactions,
the typical loop length diverges in thermodynamic limit and BP 
is therefore expected to become accurate. 

\begin{figure}[htb]
\hspace{2mm}
\includegraphics[width=0.45\columnwidth]{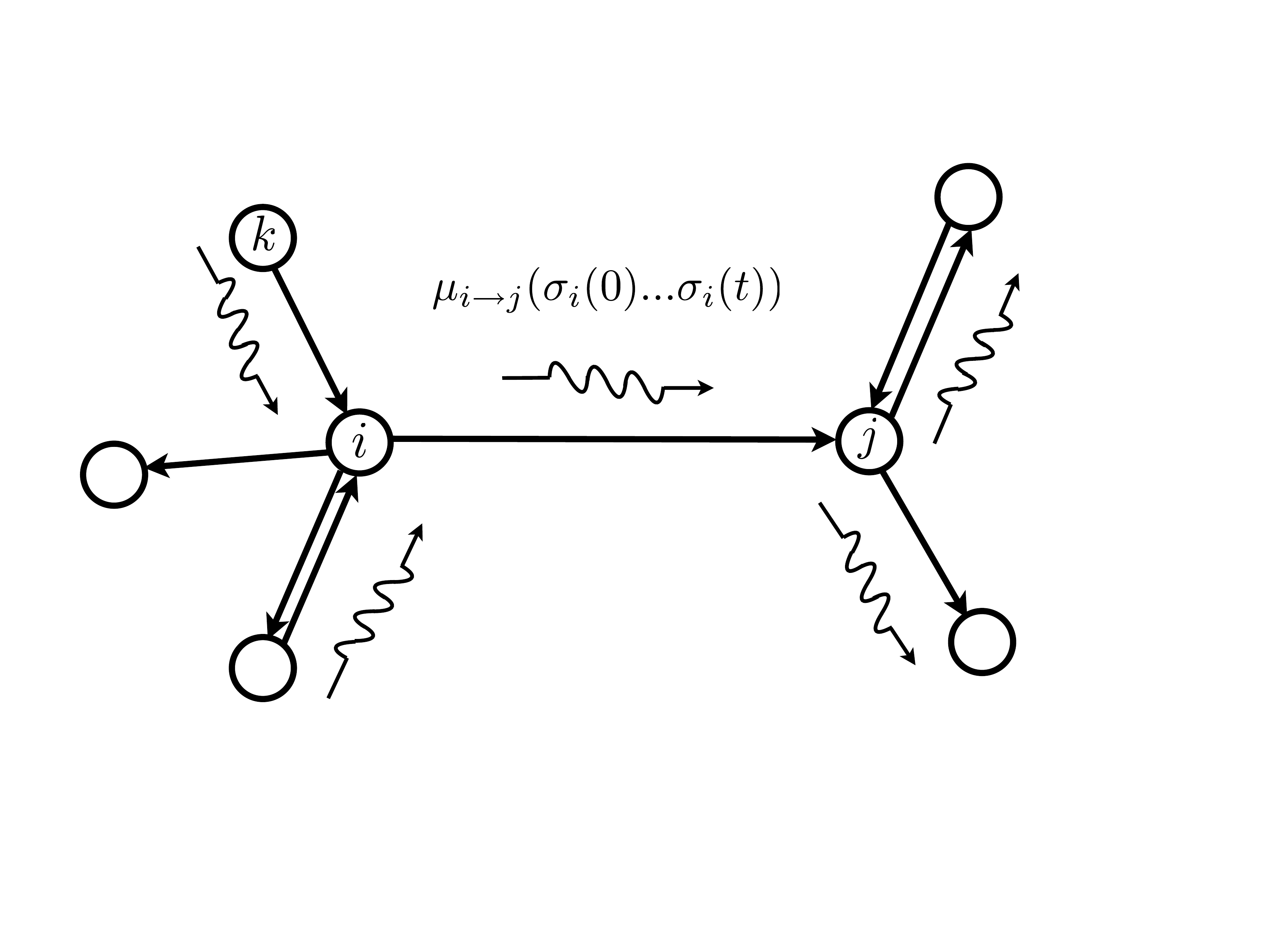}
\caption{ Dynamic message passing scheme for the diluted Ising systems. Each message
contains information describing the evolution of marginal probability in the cavity graph. Messages are
exchanged among the interacting pairs. The history of target vertex may effect the incoming message.}
\label{fig:message}
\end{figure}

By the same argument we can 
marginalize the joint probability distribution over any
single vertex in the neighborhood of $i$ and consequently all individual spins. 
The cavity assumption imposes the marginal probability $\mu_{i\to j}$
to be dependent on the spins that are directly connected to the vertex $j$.
Therefore we can interpret marginal probabilities in the cavity graph to be
a set of ``messages''  exchanged among interacting pairs. These messages 
themselves obey recursion equations (``BP update equations'')
which for parallel update are 

\begin{eqnarray}
\mu_{j\to i}(\sigma_j(0),...,\sigma_j(t)|\theta^{(i)}_j(0),...,\theta^{(i)}_j(t)) =\sum_{\sigma_{\partial j\setminus i}(0),...,\sigma_{\partial j\setminus i}(t-\Delta t)} \prod_{k\in\partial j\setminus i}&\mu_{k\to j}(\sigma_k(0),...,\sigma_k(t-\Delta t)|\theta^{(j)}_k(0),...,\theta^{(j)}_k(t-\Delta t))&\nonumber\\
&\prod_{s=1}^{t} w_j(\sigma_j(s)\,|\,h_j^{(i)}(s))\,\,\mu_{j\to i}(\sigma_j(0))&
\label{eq:bp_mess_syn}
\end{eqnarray}
Here $h_j^{(i)}$ is the effective field on spin $j$ in the cavity graph 
\begin{equation}
h_j^{(i)}(s) = \sum_{k\in\partial j\setminus i} \frac{J_{kj}}{c}\,\sigma_k(s-\Delta t) + \theta_j(t)
\label{eq:cavity-field-def}
\end{equation}
and $w_j(\sigma_j\,|\,h_j^{(i)}(s))$ is the transition probability for the single spin $j$ in the cavity graph.
Similarly for the sequential update we have (see appendix for details)
\begin{eqnarray}
\mu_{j\to i}(\sigma_j(0),...,\sigma_j(t)|\theta^{(i)}_j(0),...,\theta^{(i)}_j(t)) =&\sum_{\sigma_{\partial j\setminus i}(0),...,\sigma_{\partial j\setminus i}(t-\Delta t)} \prod_{k\in\partial j\setminus i}\mu_{k\to j}(\sigma_k(0),...,\sigma_k(t-\Delta t)|\theta^{(j)}_k(0),...,\theta^{(j)}_k(t-\Delta t))&\nonumber\\
&\prod_{s=1}^{t}\,\left[\frac{1}{N}w_j(\sigma_j(s)\,|\,h_j^{(i)}(s))+(1-\frac{1}{N})\delta_{\sigma_j(s),\sigma_j(s-\Delta t)}\right]\,\,\mu_{j\to i}(\sigma_j(0))&
\label{eq:bp_mess_asy}
\end{eqnarray}
Fig.\ref{fig:message} illustrates how messages are 
distributed among interacting vertices:
using the terminology of belief 
propagation (BP), the conditional probability $\mu_{j\to i}(\sigma_j(0,\ldots,\sigma_j(t)))$ 
can be interpreted as a message sent from variable $j$ to its neighbor $i$
 to indicate the probability of observing spin history $\sigma_j(0)\ldots\sigma_j(t)$ 
when $i$ is removed from the network.

The probability of the full history of one spin (``BP output equation'')
is on the same level of approximation
\begin{eqnarray}
p_i(\sigma_i(0),...,\sigma_i(t)\,|\,\theta_i(0),...,\theta_i(t)) = &\sum_{\sigma_{\partial i}(0)...\sigma_{\partial i}(t-\Delta t)}\,
\prod_{k\in\partial i}\mu_{k\to i}(\sigma_k(0),...,\sigma_k(t-\Delta t)|\theta^{(i)}_k(0),...,\theta^{(i)}_k(t-\Delta t))&\nonumber\\
&\prod_{s=1}^{t}\,W_i(\sigma_i(s)\,|\,h_i(s))\,\, p_i(\sigma_i(0))&
\label{eq:mar_cav}
\end{eqnarray}
A peculiarity of this formula
(shared with~Eq(\ref{eq:bp_mess_syn}) and Eq(\ref{eq:bp_mess_asy}))
is that the probability distribution of $i$ depends on the
neighbors $\partial i$ through the effective field $h_i(s)$, but the messages sent from 
the neighbors to $i$ also depend parametrically on the history of $i$ through the
modified external fields $\theta^{(i)}_k$. This difficulty is absent for fully
asymmetric networks, since then the messages sent to $i$ do not depend on the history
of $i$, as we briefly review below in Subsection~\ref{sec:full_asym}.
In the general case of partially or fully symmetric networks this difficulty is addressed 
in long time limit evolution by a stationary assumption and
the time factorization ansatz in Subsection~\ref{sec:stationary}.

Starting from a suitable initial conditions for messages, the evolution of 
messages can be followed from equation (\ref{eq:bp_mess_syn}) and (\ref{eq:bp_mess_asy}).
Each message contains $2^t$ different states corresponding to the trajectory of vertices.
Marginalizing equation (\ref{eq:bp_mess_syn}) and (\ref{eq:bp_mess_asy}) to the single time step $t$
would provide us the time dependent magnetization through equation~(\ref{eq:marg_prob})
\begin{equation}
\mu_{j\to i}^t(\sigma_j(t)|\theta_j^{(i)}(t)) = \sum_{\sigma_j(0)\ldots \sigma_j(t-1)}\mu_{j\to i}(\sigma_j(0),...,\sigma_j(t)|\theta^{(i)}_j(0),...,\theta^{(i)}_j(t)) 
\label{eq:single_message}
\end{equation}
However due to the history of target vertex $i$ the single time message does not obey a Markovian process. 
Therefore one can only hope to proceed with this
iterative procedure for only a few initial time steps. 
\subsection{Dynamic cavity method in fully asymmetric networks}
\label{sec:full_asym}
Fully asymmetric diluted Ising models where if spin $i$ connects to spin
$j$ then spin $j$ does not connect back to spin $i$ have, as alluded to
above, the simplifying property that influences (through interactions)
do not return. Accordingly, the dynamic cavity models also simplify.
In the model family considered here, $\epsilon =0$ corresponds to
the case where the probabilities of having two directed links between
spin pairs are independent. For fixed connectivity, the possibility of 
having simultaneously two links between spin pairs can be neglected in the thermodynamic limit,
this case can hence be assimilated to a fully asymmetric network.
For the dynamic cavity messages we then have
\begin{equation}
\mu_{k\to i}(\sigma_k(0),\ldots,\sigma_k(t)|\sigma_i(0),\ldots,\sigma_i(t)) = \mu_{k\to i}(\sigma_k(0),\ldots,\sigma_k(t)) \,\,\,\,\, .
\end{equation}
In parallel update this property allow us to sum over all the history except the last time step
from Eq(\ref{eq:bp_mess_syn}) and Eq(\ref{eq:single_message}), resulting in 
\begin{equation}
\mu^{t}_{i\to j}(\sigma_i(t)) = \sum_{\vec{\sigma}_{\partial i\setminus j}(t-1)}\,\prod_{k\in\partial i\setminus j} \mu^{t-1}_{k\to i}(\sigma_k(t-1))\,\,w_i(\sigma_i(s)\,|\,h_i^{(j)}(s))
\label{eq:bp_mess_syn_fully_asym}
\end{equation}
where we have introduced $\mu^t_i(\sigma_i(t))$ as the message sent from $i$ to $j$ at time step $t$.
Despite the general non-Markovian dynamics of the marginalization to one time instance in
dynamic cavity, 
 for parallel update the fully asymmetric case does follow a Markovian dynamics: at any 
 time, the messages carry information from the incoming messages
at only one time step before. Note that, resulting Markovian dynamics in this case is a consequence of update rule and asymmetry properties in the 
coupling interactions and is not necessarily restricted to belief propagation approximation. The exact dynamics in fully asymmetric case would follow
a similar equation for the joint probability distribution in which at each time the information from one time step is required. However, computationally, this is unfeasible to
study the evolution of large set of coupled spins. The case of random sequential update is more delicate, and will be discussed separately~\cite{next}.

We now consider the transition matrix to represent the Glauber dynamics
\begin{equation}
w_i(\sigma_i(s)\,|\,h_i(s)) = \frac{e^{\beta\,h_i(s)\,\sigma_i(s)}}{2\,\cosh(\beta\,h_i(s))}
\end{equation}
from which the single site magnetization under parallel updates follows
\begin{equation}
m_i(t) = \sum_{\vec{\sigma}_{\partial i}(t-1)} \prod_{j\in \partial i}\,\mu^t_{j\to i}(\sigma_j(t-1))\,\tanh\left[\beta(\sum_{j\in\partial i}J_{ji}\,\sigma_j(t-1)+\theta_i)\right] \,\,\,\,\, .
\end{equation}
Due to the particular form of transition matrix in the Glauber dynamics (local normalization),
the resulting dynamic BP equations in the fully asymmetric networks yet requires 
a summation over the whole configuration of neighboring vertices. Note that
some more simplified transition matrices would provide us with an even more efficient BP approximation 
where the number of required summations scales linearly with the size of neighboring vertices\cite{montanari}. 

As an interesting extension, the fully connected graphs with weak interactions can be realized by taking
the limits $c\to N$ and $N\to \infty$. Since the interaction couplings are scaled with 
$1/c$ the variables become weakly connected in this limit and the graph 
statistically uncorrelated. 
Introducing $\delta h_i(t) = \sum_{j\in\partial i}\frac{J_{ji}}{c}(\sigma_j(t-1)-m_j(t-1))$ 
we can expand the right hand side around $\delta h_i$ considering 
the fluctuation of spins to be small with respect to their mean value at each time
\begin{equation}
m_i(t) = \tanh\left[\beta(\sum_{j\in\partial i}\frac{J_{ji}}{c}\,m_j(t-1)+\theta_i)\right]+{\cal O}(c^{-2})
\end{equation}
To first order in $1/c$ we end up with the same equation 
for naive mean field approximation as the one introduced by~\cite{derrida}
and to the second order in $1/c$ with the dynamic TAP approximation~\cite{yasser_tap,kappen}.

\subsection{Dynamic cavity method and stationary states}
\label{sec:stationary}
In this subsection we introduce an approximation scheme for general
(not fully asymmetric) networks assuming that the system is in a stationary
state. 
The problem to be solved is the non-Markovian nature of the 
evolution of the single-spin single-time marginals
(\textit{e.g.} Eq(\ref{eq:single_message}))
which follows from marginalization of the full dynamic cavity equations
(\ref{eq:bp_mess_syn}) and (\ref{eq:bp_mess_asy}) over time.
The approximation is that the dynamic cavity messages factorize over time:
\begin{equation}
\mu_{i\to j}(\sigma_i(0),\ldots,\sigma_i(t)\,|\,\sigma_j(0),\ldots,\sigma_j(t))  = \prod_{s=0}^t\,\, \mu^t_{i\to j}(\sigma_i(s)\,|\,\sigma_j(s-1))
\end{equation} 
This {\it time-factorization} assumption is clearly
not appropriate to describe transients, where we would expect both dependencies
between messages and that the functional form of the messages depend on time.
However, in a stationary state it may be acceptable to take the messages independent
in time, and it is reasonable to assume that the single-time messages do not depend
explicitly on time. 
On the other hand, from the computational point of view, time-factorization
provides us a closed set of equations for the single-time marginals, which 
makes the whole scheme computationally feasible: 
\begin{equation}
\mu^t_{i\to j}(\sigma_i(t)) = \sum_{\sigma_{i}(t-2),\vec{\sigma}_{\partial i\setminus j}(t-1)}\, \prod_{k\in\partial i\setminus j}\,\mu^{t-1}_{k\to i}(\sigma_k(t-1)|\sigma_i(t-2))\,w_i(\sigma_i(s)\,|\,h_i^{(j)}(s))\,\,\mu^{t-2}_{i\to j}(\sigma_i(t-2))
\label{eq:stationary_mess}
\end{equation} 
In above, for parallel updates, as treated in~\cite{bolle},
the summation over time history is resulted in $\mu_{i\to j}^{t-2}(\sigma_i(t-2))$.
We refer to appendix for the corresponding equations for sequential updates.
Note that in this approximation the
single-time dynamic cavity messages at time $t$ depend on messages sent at 
 most two time steps earlier.  For Ising spins, we can write the single-time dynamic 
cavity messages using cavity biases
$\mu^t_{i\to j}(\sigma_i(t)) = \frac{\beta\,u_{i\to j}(t)\,\sigma_i(t)}{2\cosh(u_{i\to j}(t))}$,
and inserting this equation into~(\ref{eq:stationary_mess}) we get 
an evolution equation for the cavity biases 
\begin{eqnarray}
u_{i\to j}(t) = \frac{1}{2\beta}\sum_{\sigma_i(t)}\,\sigma_i(t)\log\left[\sum_{\vec{\sigma}_{\partial i\setminus j}(t-1),\sigma_i(t-2)}\frac{e^{\beta \left(\sum_{k\partial i\setminus j}u_{k\to i}+J_{ik}\sigma_i(t-2)\right)\sigma_k(t-1)}}{\prod_{{k\partial i\setminus j}}2\cosh[\beta \left(u_{k\to i}+J_{ik}\sigma_i(t-2)\right)]}\right.\nonumber\\
\left.\frac{e^{\beta\,h_i^{(j)}(t)\sigma_i(t)}}{2\cosh(\beta\,h_i^{(j)}(t))}\,\,\frac{e^{\beta u_{i\to j}(t-2)\sigma_i(t-2)}}{2\cosh(\beta u_{i\to j}(t-2))}\right]
\label{eq:cavity-bias-eq}
\end{eqnarray}
supplemented by Eq(\ref{eq:cavity-field-def})
\textit{i.e.}
$h_j^{(i)}(t) = \sum_{k\in\partial j\setminus i} \frac{J_{kj}}{c}\,\sigma_k(t-1) + \theta_j(t)$
for the cavity fields.
Solving for the stationary state of the kinetic Ising model using dynamic cavity
equations in the time-factorized approximation hence means to find a
fixed point of (\ref{eq:cavity-bias-eq}) and (\ref{eq:cavity-field-def}) when the 
external fields $\vec{\theta}$ are independent
in time.
Note that $\sigma_i(t-2)$ contributes only when the edges from $i$ to $j$ and
from $j$ to $i$ are both present.
Therefore, in the fully asymmetric network this term disappears and we get back to 
Eq(\ref{eq:bp_mess_syn_fully_asym}). 
On the other hand, for fully symmetric networks, as has been already pointed out in~\cite{bolle} and also in GFA analysis in~\cite{coolen1},
the ordinary belief propagation equation is a solution of dynamic cavity equation in the time factorized approximation.
Indeed, it can be verified that Eq(~\ref{eq:cavity-bias-eq}) for $c_{ij}J_{ij}=c_{ji}J_{ji}$ admits a solution of the form $u_{i\to j} = \theta_i + 1/\beta \sum_{k\in \partial i\neq j}\tanh(\beta J_{ki}\tanh(u_{k\to i}))$ 
that is the ordinary belief propagation equation for Ising systems with pairwise interactions.
We note also that in a transient we can
compute the time evolution of magnetization from Eq(\ref{eq:marginal})
\begin{eqnarray}
m_i(t) = \sum_{\vec{\sigma}_{\partial i\setminus j}(t-1),\sigma_i(t-2)}\,\frac{e^{\beta \left(\sum_{k\partial i\setminus j}u_{k\to i}+J_{ik}\sigma_i(t-2)\right)\sigma_k(t-1)}}{\prod_{{k\partial i\setminus j}}2\cosh[\beta \left(u_{k\to i}+J_{ik}\sigma_i(t-2)\right)]}\nonumber\\
\tanh\left[\beta(\sum_{j\in\partial i}J_{ji}\,\sigma_j(t-1)+\theta_i)\right]\,\frac{e^{\beta u_{i\to j}(t-2)\sigma_i(t-2)}}{2\cosh(\beta u_{i\to j}(t-2))}\
\end{eqnarray}
This is not expected to be accurate unless we are already in a stationary state,
but will be used below in Section~\ref{sec:result} to monitor the approach
to the stationary state.
\section{Results}
\label{sec:result}
In this section we investigate the stationary states of diluted spin glass 
by dynamic cavity approach in both parallel and sequential update. 
We show how the total
magnetization of diluted Ising systems as computed by dynamic cavity method evolves with time. 
In order to verify the results, we 
perform a numerical simulations (Monte Carlo) based on the appropriate dynamics.

\subsection{BP versus Glauber dynamics}
\begin{figure}[htb]
\vspace{0.6cm}
\includegraphics[width=0.31\columnwidth]{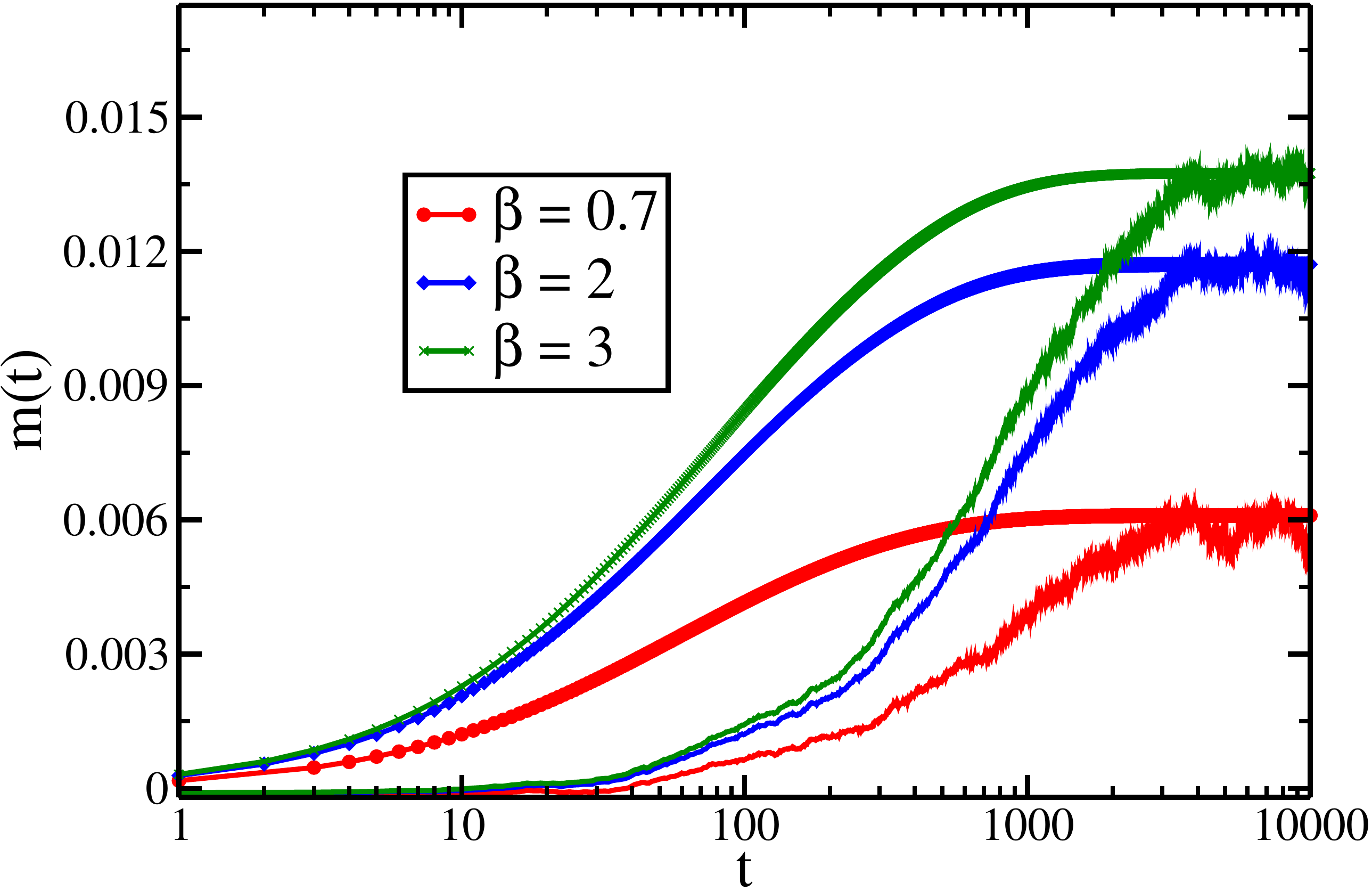}
\hspace{2mm}
\includegraphics[width=0.31\columnwidth]{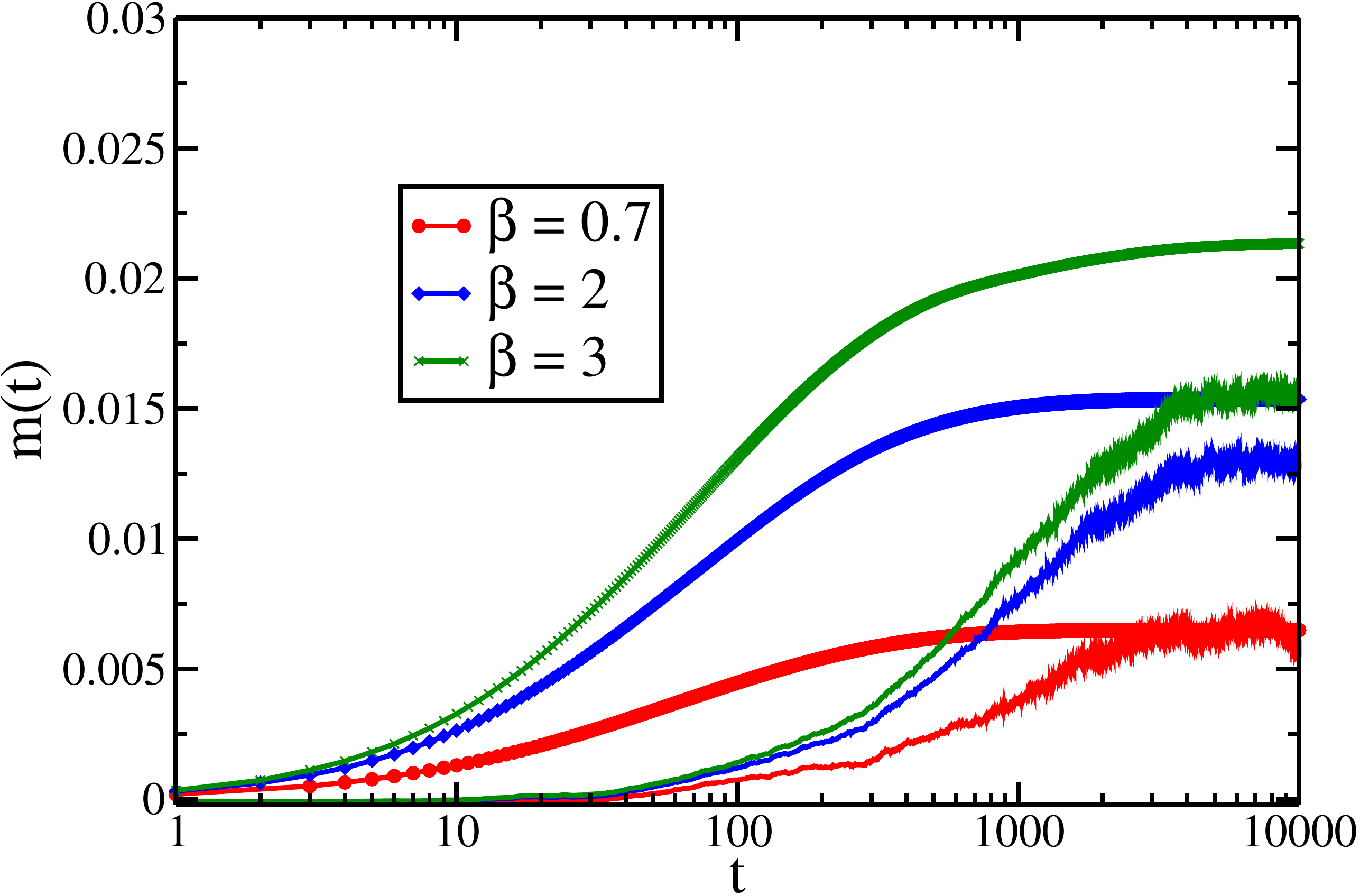}
\hspace{2mm}
\includegraphics[width=0.31\columnwidth]{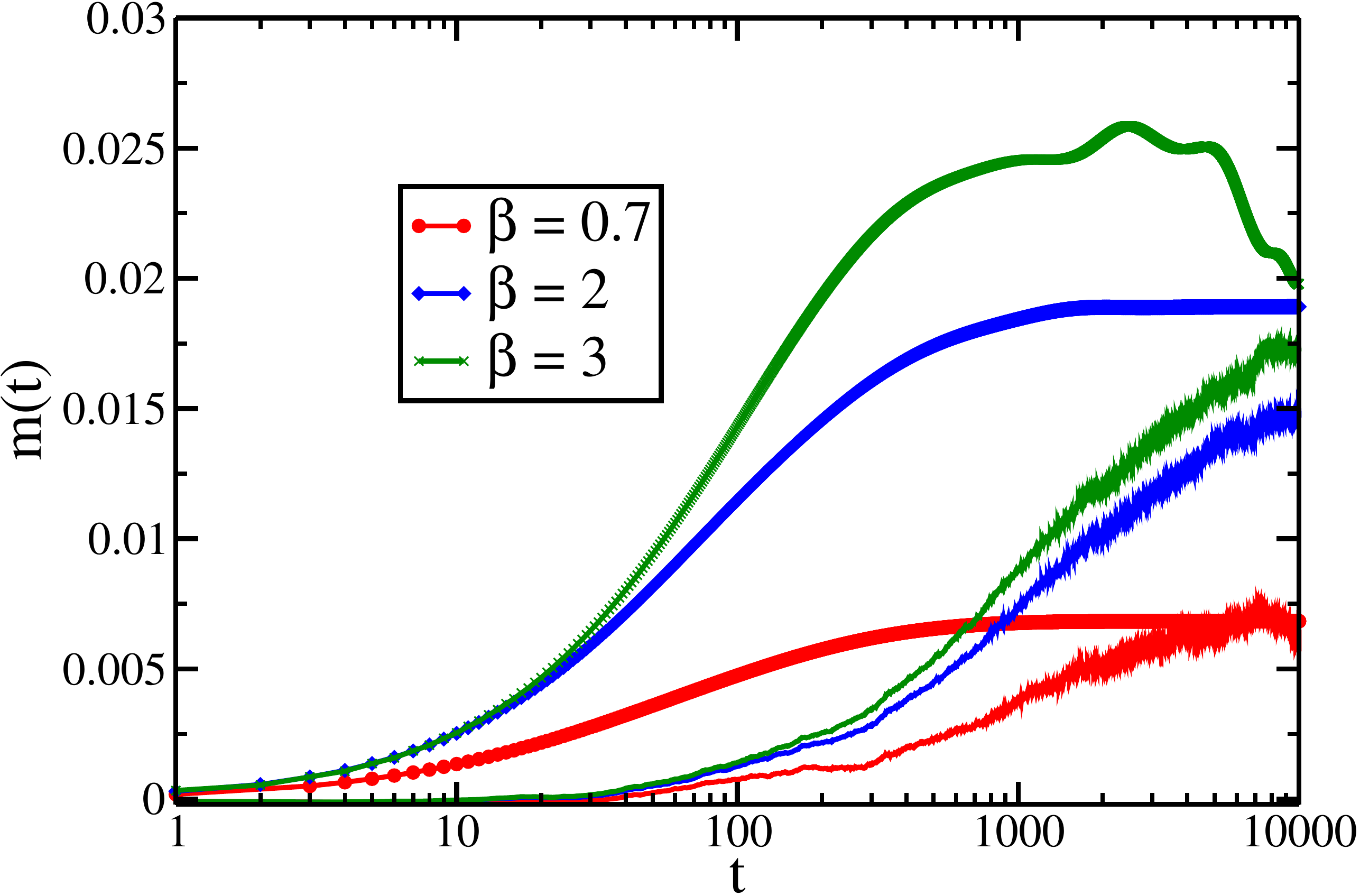}
\caption{Time evolution of magnetization in a diluted networks under the sequential update.
The interactions are random variables from normal distribution, connectivity parameter is fixed to $c=3$ and simulations are
performed for $\epsilon = 0, 0.5, 1$ and $\beta = 0.7, 2, 3$. A non-zero static external fields $\theta_i=0.01$
acts on each individual spin.
Left panel, fully asymmetric network, middle panel partially asymmetric with $\epsilon=0.5$, right panel 
a fully symmetric network. Symbols show the dynamic cavity method and
solid lines show the numerical simulations by Monte Carlo averaged over $10^4$ samples of the size $N = 1000$. }
\label{fig:evol_sequential}
\end{figure}

We first generate diluted graphs from random ensemble
 of size $10^3$ and $10^4$ using asymmetry and connectivity parameters as in Eq(\ref{eq:connec}) and Eq(\ref{eq:asym}).
 For each graph instance we iterate the dynamic cavity equations and simulate Monte Carlo analysis
 to test the accuracy of dynamic cavity method. The system is initialized to a random configuration
 with small external fields acting on each individual variable.
 In Monte Carlo simulations, we generate up to $10000$ samples to 
 estimate the time evolution of magnetization at short time steps.
We focus on the high-temperature regime
to avoid the spin glass phase (for networks with symmetric interaction couplings,
and presumably also for weakly asymmetric networks).
Typical results for the time evolution of total magnetization $m(t)=1/N\sum_{i=1}^N m_i(t)$ under sequential updates are
illustrated in Fig.\ref{fig:evol_sequential}.
For short times, dynamic cavity method differs significantly from Monte Carlo, and
also displays its own (unrelated) dynamics. For long times however dynamic cavity
reaches a fixed point, and this stationary state agrees well with Monte Carlo
(for the magnetizations).
The time needed for Monte Carlo to reach the stationary state grows with $\beta$
as it can be seen from the Fig.\ref{fig:evol_sequential}, while for dynamic cavity in sequential update
there is no such noticeable dependence except for the fully symmetric networks
(right panel).

\begin{figure}[htb]
\vspace{0.6cm}
\includegraphics[width=0.45\columnwidth]{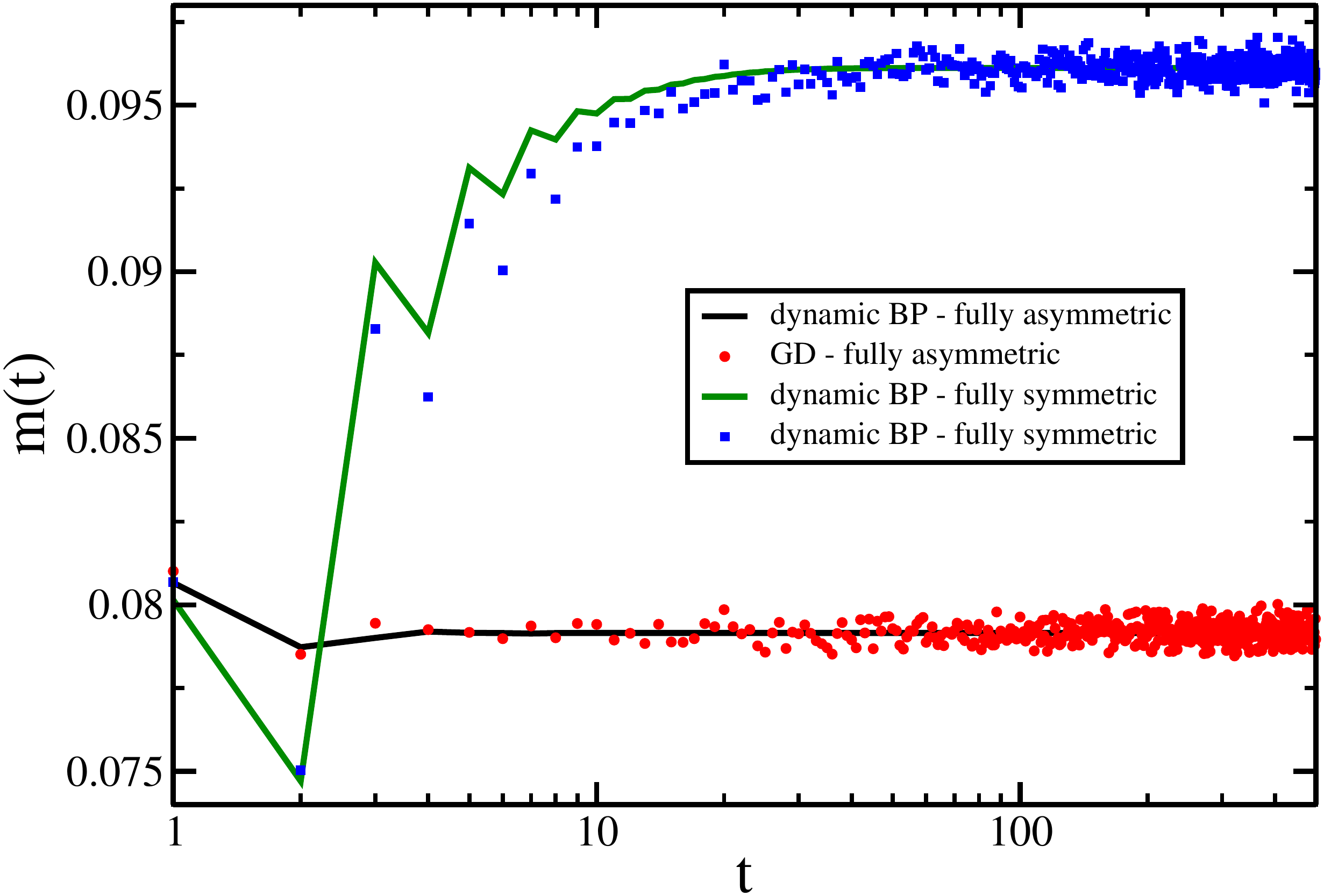}
\hspace{3mm}
\includegraphics[width=0.45\columnwidth]{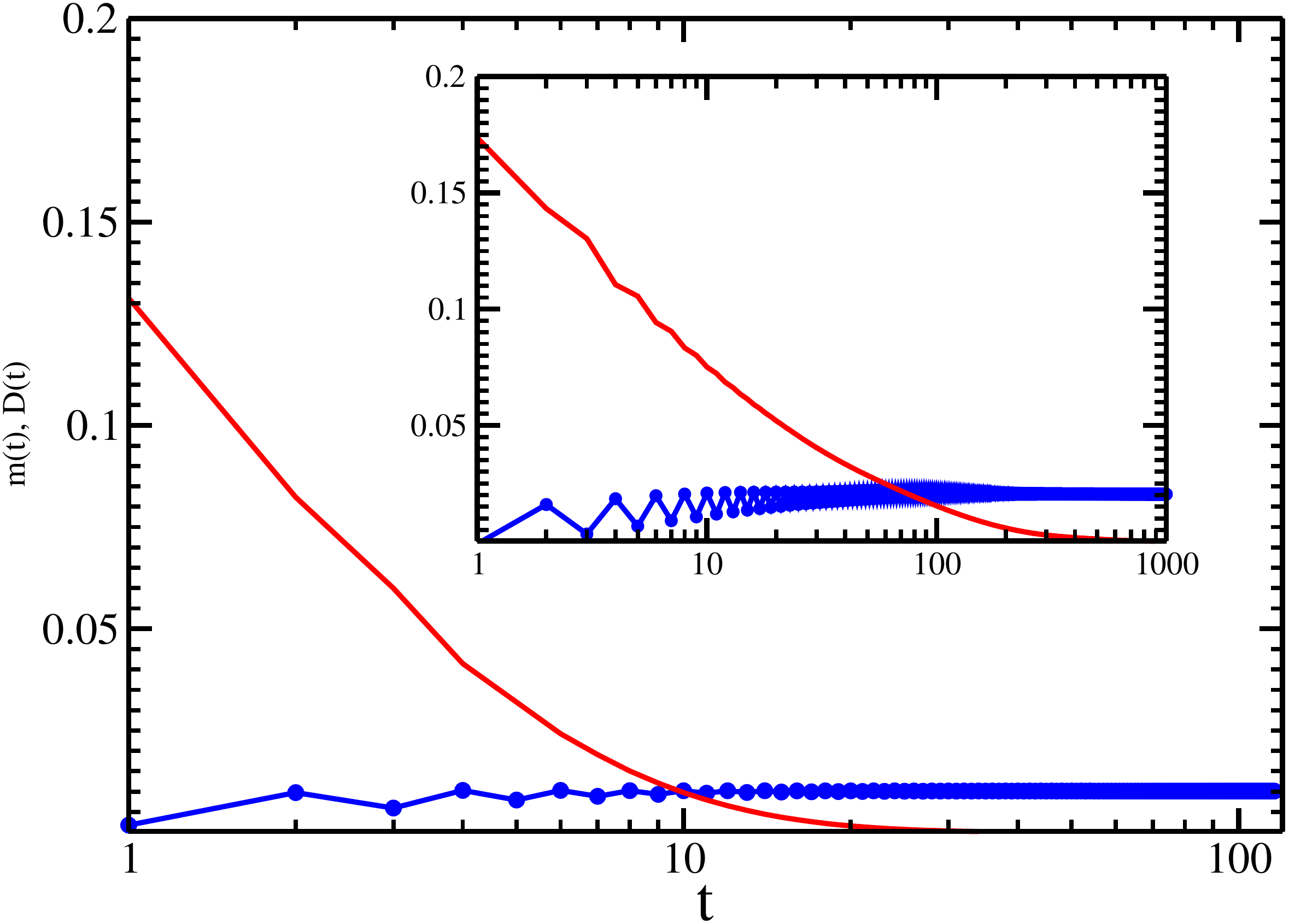}
\caption{Left panel : the time evolution of total magnetization at $\beta=1$
for networks of size $N=10^4$ with asymmetric and fully symmetric interactions.
 In all cases, a stationary solution according to dynamic BP exist which is consistent with numerical simulations. 
 Right panel:  the comparison between fixed point of ordinary BP and the time evolution of magnetization obtained 
 by dynamic cavity method for fully symmetric networks. Inverse temperature $\beta$ is chosen to be $1$ and $2$.
 In both cases, ordinary BP converges and the limit $D(t)\to 0$ exists. For lower $\beta$ a faster convergence of 
 dynamic cavity method is found.   
}
\label{fig:evol}
\end{figure}

For parallel dynamics, the situation is more involved. 
For low enough $\beta$, again we have consistent results for dynamic cavity and Monte Carlo simulations (Fig.~\ref{fig:evol}, left panel).
Note that the dynamic cavity method for diluted fully asymmetric network in parallel update is exact, and (Fig.~\ref{fig:evol}, left panel) lower curve gives a measure of the numerical fluctuations.   
In (Fig.~\ref{fig:evol}, right panel) we verify that for symmetric interactions and in low enough $\beta$, the solution for dynamic 
cavity is also a solution of ordinary BP. 
Introducing $D(t) = 1/N\sum_{i=1}^N(m_i(t)-m_i^{(BP)})^2$ 
where $m_i^{(BP)}$ is the single magnetization obtained from output belief propagation (equilibrium), we
expect to get a zero value for $D(t)$ by evolving dynamic cavity during time. Fig.~\ref{fig:evol}, right panel, shows the evolution of $D(t)$
for $\beta=1$ and $\beta =2$. For both cases, ordinary BP converges to a fixed point and dynamic cavity method predicts
stationary solution to the time dependent magnetization. For $\beta=1$ a fast convergence for dynamic cavity solution is observed to the 
total magnetization predicted by ordinary BP. Increasing $\beta$ (still in the phase where ordinary BP converges) would require longer 
relaxation time for constant results.  

In the large $\beta$ limit, however, the 
system may fall into limit cycles with no stable stationary state. This can be observed by 
computing total magnetization over long periods of time. 
In Fig.\ref{fig:delta} the time evolution of total magnetization in a fully symmetric networks computed by dynamic cavity is illustrated for 
$\beta = 4$ and $\beta= 5$.
We observe that even in long time limit, the dynamics according to cavity method, does not approach a fixed solution but 
roughly oscillates between two states.
Interestingly, for fully asymmetric networks we do not observe such cyclic behavior.
\subsection{Cyclic stationary states}
\begin{figure}[htb]
\vspace{0.6cm}
\includegraphics[width=0.45\columnwidth]{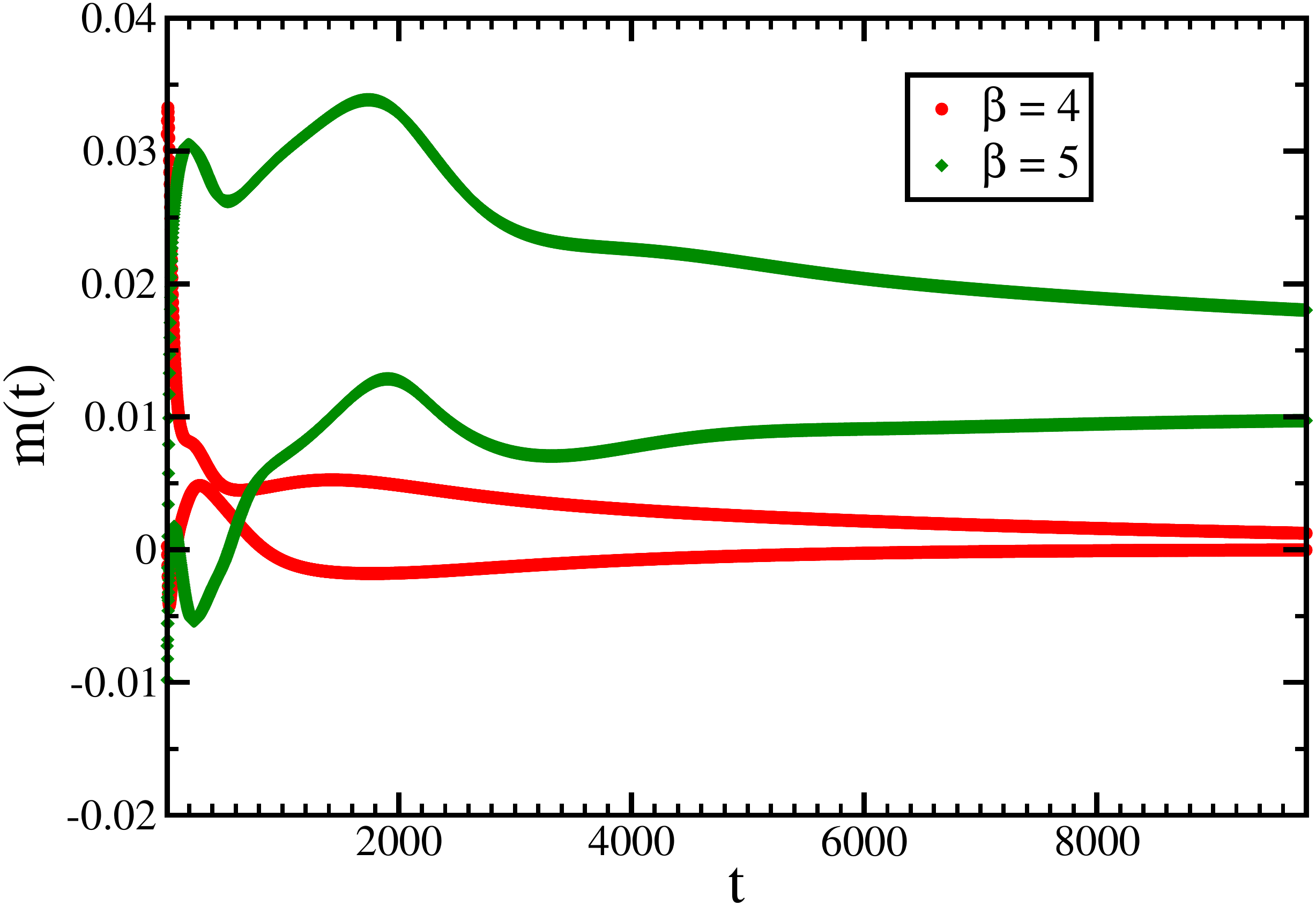}
\hspace{3mm}
\includegraphics[width=0.45\columnwidth]{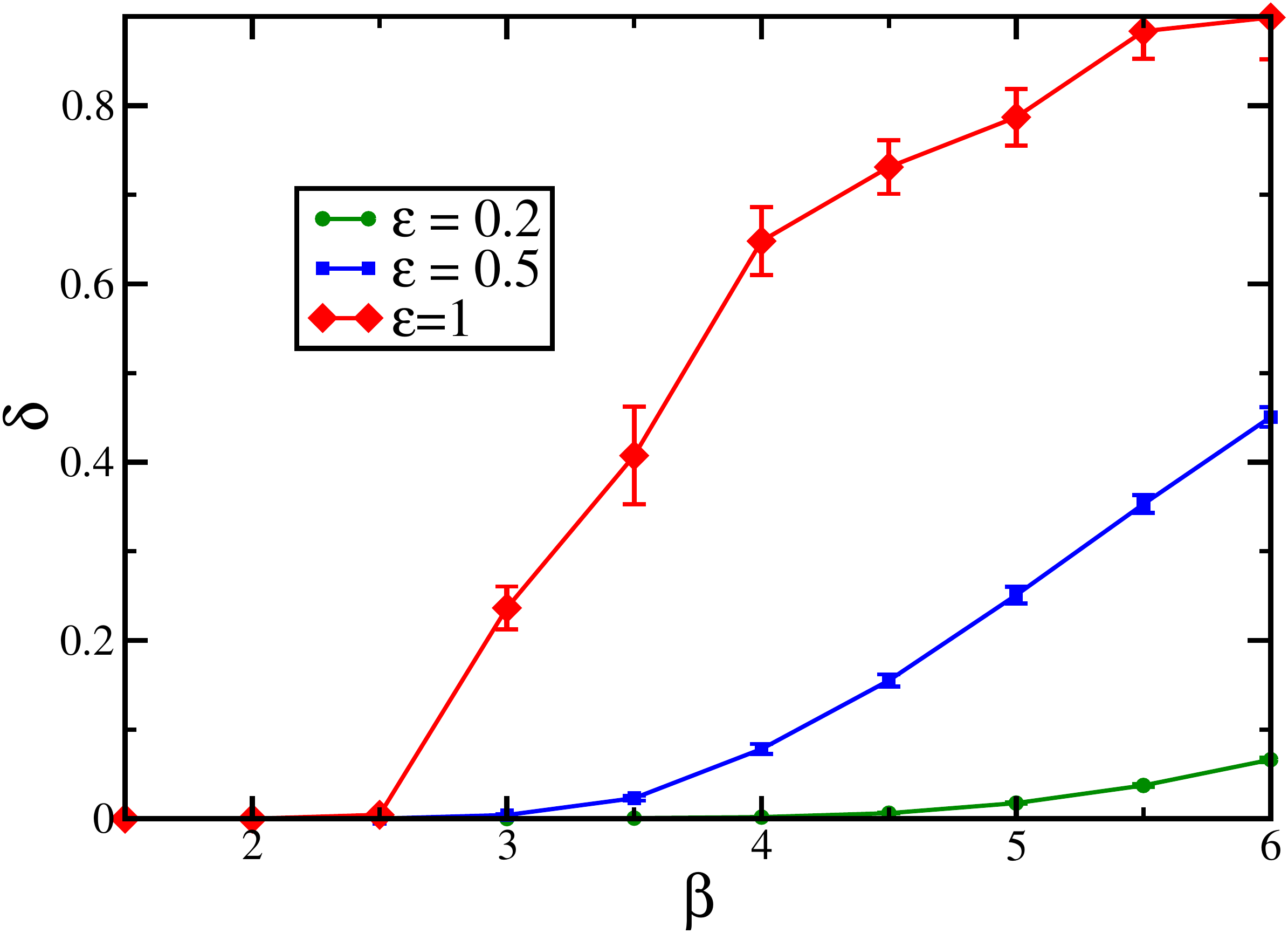}
\caption{Dynamic cavity results for diluted spin glass systems in parallel update at low temperature.
The interactions are random variables from normal distribution, the system size is $N=1000$ with an average connectivity $c=3$. 
Left panel, the time evolution of magnetization for $\epsilon=0.5$ and $\beta=4,5$ in absent of external fields. It shows an oscillatory behavior 
even at long limit time (for non-fully asymmetric and low temperature). 
Right panel, the evolution of $\delta$ for system size of $N=1000$ after $t=10^4$ steps and for $\epsilon = 0.2, 0.5, 1$. For fully
asymmetric networks $\delta =0$ showing the existence of stable stationary state. }
\label{fig:delta}
\end{figure}

In order to study the cyclic behavior of stationary states in parallel update we introduce
a time dependent quantity which measures the difference of total magnetizations in two successive time steps 
\begin{equation}
\delta(t) = \frac{1}{N}\sum_{i=1}^N\, (m_i(t)-m_i(t-1))^2
\end{equation}
A zero value for $\delta(t)$ indicates the existence of stationary state.
On the other hand, non-zero $\delta(t)$, even after long times, means that either
a stationary state does not exist, or is not reachable in finite time.
Our results (from Fig.~\ref{fig:delta} and data not shown)
are that cyclic behavior (of this type), is observed for
parallel updates at sufficiently low temperature if the network is not fully
asymmetric, but not for sequential updates. The strongest cyclic behavior belongs to
fully symmetric networks. At fixed connectivity
the amplitude of the oscillations tend to decrease with system size (data not shown).
\section{Conclusion and Discussion}
\label{sec:conclusion}
In this paper we have studied stationary states of diluted 
asymmetric (kinetic) Ising models by the dynamic cavity method,
both for parallel updates (as in~\cite{bolle} and \cite{montanari}) and for
sequential updates (new). We find that for a many such systems with
different asymmetry, sparseness and interaction strengths, total
magnetization computed by dynamic cavity matches direct numerical
simulation (Monte Carlo) -- at many orders of magnitude less 
computational effort.

Nothing does however come for free, and for all cases except a fully
asymmetric network in parallel update, the dynamic cavity method presumably fails
at low enough temperature. We observe (for the cases we have studied)
different failure modes for
different update rules, where for parallel updates the method simply
no longer converges (no longer finds a stationary state), while for 
sequential updates the method converges to fixed point, but which does
not correspond to the stationary state found by direct simulation.

The full phase diagram of these models remains to be done; we have \textit{e.g.}
not verified that dynamic cavity fails at low temperature for \textit{all}
models except fully asymmetric ones (although we expect that to be true).
The comparison should also be done in a more detailed spin-by-spin
manner, and extended at least to pair-wise correlations. We intend
to return to these topics in a future contribution.

The sequential update model is close to an asynchronous update model (continuous time)
\textit{i.e.} the master equation, as simulated \textit{e.g.} by the
Gillespie algorithm. But at least for finite $N$, the two models are
not identical, and the analysis carried out here should
be extended to the master equation (if this is possible).

Finally, since kinetic Ising models have recently been used for
inference and dynamic network reconstruction~\cite{erik,yasserhertz}, and
since ordinary BP has been used for inference in equilibrium systems~\cite{mezard-mora}
it would be interesting if the two strands could be combined. 

\section*{\it Acknowledgment}
We thank Dr. Izaak Neri for interesting discussions
and Institute of Theoretical Physics - Chinese Academy of Sciences
(Beijing, China) for hospitality. The work was supported by the
Academy of Finland as part of its Finland Distinguished
Professor program, project 129024/Aurell.
\subsection{Appendix}
The dynamic cavity approach to the parallel update was first derived by Neri and Bolle in ~\cite{bolle}.
Here we show how it can be extended to sequential update. In the sequential update
we assume that at each time step, one single variable is selected randomly and will be updated according to Eq(~\ref{eq:dynamic}).
The time evolution of probability distribution of all variables, follows
\begin{equation}
p(\vec{\sigma}(0),\ldots,\vec{\sigma}(t)) = \prod_{s=1}^t\, W(\vec{\sigma}(s)\,|\,\vec{h}(s))\, p(\vec{\sigma}(0))
\end{equation} 
where the transition probability contains the sequential update. 
The appropriate choice for the single variable update is 
the uniform probability distribution over all spins (see \cite{coolen_introduction})
\begin{equation}
W(\vec{\sigma}(t)\,|\,\vec{h}(t)) = \frac{1}{N} \sum_{i=1}^N\, \left\{\prod_{j\neq i}\delta_{\sigma_j(t),\sigma_j(t-\Delta t)}\, w_i(\sigma_i(t)\,|\,h_i(t))\right\}
\end{equation}
Following the cavity graph argument discussed in section \ref{sec:BP}, the evolution of marginal probability follows 
\begin{eqnarray}
p_i(\sigma_i(0),\ldots,\sigma_i(t)) = \frac{1}{N}\sum_{\vec{\sigma}_{\partial i}(0),..,\vec{\sigma}_{\partial i}(t)}\,  \prod_{k\in\partial i}\,\mu_{k\to i}(\sigma_k(0),\ldots,\sigma_k(t-\Delta t)|\theta^{(i)}_k(0),\ldots,\theta^{(i)}_k(t-\Delta t))\nonumber\\
 \prod_{s=\Delta t}^t \sum_{j=1}^N\, \left\{\prod_{k\neq j}\delta_{\sigma_k(s),\sigma_k(s-\Delta t)}\, w_i(\sigma_j(s)\,|\,h_j(s))\right\}\, p_i(\sigma_i(0))
\label{eq:marg_sequential}
\end{eqnarray}
The summation over variable update inside the formula will contribute in two terms: either the index $j$ is equal to the cavity variable $i$
or is different.
We can now introduce messages among neighboring variables carrying the marginal probability in the cavity graph. They fulfill the same 
recursive equation as parallel update
 \begin{eqnarray}
\mu_{i\to j}(\sigma_i(0),\ldots,\sigma_i(t)|\theta^{(j)}_i(0),\ldots,\theta^{(j)}_i(t)) &=&\sum_{\vec{\sigma}_{\partial i\setminus j}(0),\ldots,\vec{\sigma}_{\partial i\setminus j}(t)} \prod_{k\in\partial i\setminus j}\,\mu_{k\to i}(\sigma_k(0),\ldots,\sigma_k(t-\delta t)|\theta^{(i)}_k(0),\ldots,\theta^{(i)}_k(t-\Delta t))\nonumber\\
&&\prod_{s=\Delta t}^{t}\,\left[\frac{1}{N}w_i(\sigma_i(s)\,|\,h_i(s))+(1-\frac{1}{N})\delta_{\sigma_i(s),\sigma_i(s-\Delta t)}\right]\,\,\mu_{i\to j}(\sigma_i(0))
\end{eqnarray}
which is Eq(\ref{eq:bp_mess_asy}) used in the section(\ref{sec:BP}) for dynamic cavity in the sequential update. 
In the stationary state where the initial condition is assumed to be irrelevant we can perform the time-factorized approximation over the
time evolution of messages. The time evolution of messages at each time then simplifies following a Markovian dynamics of length two.
\begin{eqnarray}
\mu^t_{i\to j}(\sigma_i(t)|\theta_i^{(j)}(t)) =\frac{1}{N} \mu^{t-\Delta t}_{i\to j}(\sigma_i(t)|\theta_i^{(j)}(t-\Delta t)) + (1-\frac{1}{N}) \sum_{\sigma_i(t-2\Delta t),\vec{\sigma}_{\partial i\setminus j}(t-\Delta t)} \prod_{k\in\partial i\setminus j}\,\mu^{t-\Delta t}_{k\to i}(\sigma_k(t-\Delta t)\,|\,\theta_i^{(j)}(t-2\Delta t))\nonumber\\
w_i(\sigma_i(t)\,|\,h_i(t))\,\mu^{t-2\Delta t}_{i\to j}(\sigma_i(t-2\Delta t))
\end{eqnarray}

\end{document}